\documentclass[pdftex,twocolumn,3]{jour3}          

\usepackage[utf8]{inputenc}
\usepackage{epsf}
\usepackage{latexsym,amssymb,euscript}
\usepackage{widetext}
\usepackage[dvips]{graphicx}
\usepackage[numbers,sort&compress]{natbib}
\usepackage{amsmath}
\usepackage{nicefrac}
\usepackage{slashed}
\usepackage{booktabs}
\usepackage{hyperref}
\usepackage{braket}
\usepackage{chngcntr}
\usepackage{bm}
\usepackage{bbold}
\usepackage{graphics}
\usepackage{graphicx}
\usepackage{mciteplus}
\usepackage{bbold}
\usepackage{pdfpages}
\usepackage[titletoc]{appendix}
\graphicspath{{./figures/}}
\hypersetup{
 linktocpage = true,
 urlcolor = urlblue,
 colorlinks = true,
 linkcolor = urlblue,
 anchorcolor = urlblue,
 citecolor = citegreen,
 pdfstartview = {XYZ null null 1.25} 
           }
\usepackage[left=2cm, right=2cm]{geometry}
\usepackage{pstricks}
\usepackage{color}
\usepackage{xcolor}
\definecolor{urlblue}{rgb}{0.2,0.4,0.7}
\definecolor{citegreen}{rgb}{0,0.4,0.2}
\definecolor{linkred}{rgb}{0.9,0.2,0.1}
\usepackage{float}
\usepackage{academicons}
\definecolor{orcidlogocol}{HTML}{A6CE39}
\usepackage{fancyhdr}
\newcommand{\drv}{{\rm d}}

\newcommand{\tcite}[1]{~\cite{#1}}
\newcommand{\tref}[1]{~\ref{#1}}
\newcommand{\eref}[1]{~\eqref{#1}}

\newcommand{\tarr}{
\begin{array}}
\newcommand{\earr}{\end{array}}

\newcommand{\orcidAB}{\href{https://orcid.org/0000-0002-8824-8355}{\includegraphics[scale=0.1]{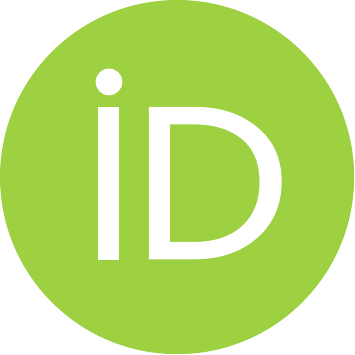}}}

\newcommand{\orcidFGC}{\href{https://orcid.org/0000-0003-3299-2203}{\includegraphics[scale=0.1]{figures/logo-orcid.pdf}}}

\newcommand{\orcidMR}{\href{https://orcid.org/0000-0002-4542-9797}{\includegraphics[scale=0.1]{figures/logo-orcid.pdf}}}

\newcommand{\orcidAS}{\href{https://orcid.org/0000-0001-6640-9659}{\includegraphics[scale=0.1]{figures/logo-orcid.pdf}}}

\smartqed  

\journalname{}

\begin{document}

\title{Phenomenology of gluon TMDs from $\eta_{b,c}$ production
}

\subtitle{}

\author{
Alessandro Bacchetta
\thanksref{e1,addr1,addr2} \orcidAB
\and
Francesco Giovanni Celiberto
\thanksref{e2,addr3,addr4,addr5} \orcidFGC
\and
Marco Radici
\thanksref{e4,addr2} \orcidMR
\and
Andrea Signori
\thanksref{e5,addr7,addr8} \orcidAS
}

\thankstext{e1}{{\it e-mail}:
\href{mailto:alessandro.bacchetta@unipv.it}{alessandro.bacchetta@unipv.it}}
\thankstext{e2}{{\it e-mail}:
\href{mailto:fceliberto@ectstar.eu}{fceliberto@ectstar.eu} (corresponding author)}
\thankstext{e4}{{\it e-mail}:
\href{mailto:marco.radici@pv.infn.it}{marco.radici@pv.infn.it}}
\thankstext{e5}{{\it e-mail}:
\href{mailto:andrea.signori@unito.it }{andrea.signori@unito.it }}

\institute{Dipartimento di Fisica, Universit\`a di Pavia, I-27100 Pavia, Italy\label{addr1}
\and
Istituto Nazionale di Fisica Nucleare, Sezione di Pavia, I-27100 Pavia, Italy\label{addr2}
\and
European Centre for Theoretical Studies in Nuclear Physics and Related Areas (ECT*), I-38123 Villazzano, Trento, Italy\label{addr3}
\and
Fondazione Bruno Kessler (FBK),
I-38123 Povo, Trento, Italy\label{addr4}
\and
INFN-TIFPA Trento Institute of Fundamental Physics and Applications,
I-38123 Povo, Trento, Italy\label{addr5}
\and
Dipartimento di Fisica, Universit\`a di Torino, I-10125 Torino, Italy\label{addr7}
\and
Istituto Nazionale di Fisica Nucleare, Sezione di Torino, I-10125 Torino, Italy\label{addr8}
}

\date{\today}

\maketitle

\section*{Abstract}

We present the potential of $\eta_{b,c}$ production in proton collisions to access the gluon transverse-momentum dependent parton distributions (TMDs). In particular, we explore the phenomenology of the unpolarized and linearly-polarized gluon TMDs in unpolarized collisions for different kinematic settings, and the potential of a fixed-target experiment at the LHC.
\vspace{0.50cm} \hrule
\vspace{0.50cm}
{
 \setlength{\parindent}{0pt}
 \textsc{Keywords}: \vspace{0.15cm} \\ QCD phenomenology \\ TMD factorization \\ Quarkonium production
}
\vspace{0.50cm} \hrule


\section{Gluon TMDs and quarkonia}
\label{introduction}

Unveiling the inner dynamics of gluons and quarks inside nucleons is a frontier research of core studies at new-generation colliding machines\tcite{Chapon:2020heu,Anchordoqui:2021ghd,Feng:2022inv,Hentschinski:2022xnd,Boer:2011fh,Accardi:2012qut,AbdulKhalek:2021gbh,Khalek:2022bzd,Acosta:2022ejc,AlexanderAryshev:2022pkx,Celiberto:2016yek,Celiberto:2016jse,Brunner:2022usy,Arbuzov:2020cqg,Abazov:2021hku,Bernardi:2022hny,Amoroso:2022eow,Celiberto:2018hdy,Klein:2020nvu,2064676,MuonCollider:2022xlm,Aime:2022flm}.
Finding an answer to QCD fundamental questions, as the origin of proton spin and mass, depends on our ability of reconstructing the three-dimensional motion of partons inside parent hadrons. This requires going beyond the standard approach based on collinear parton density functions (PDFs)~\cite{NAP25171,Accardi:2012qut,Geesaman:2015fha,AbdulKhalek:2021gbh}.
Accounting for the distribution of partons in momentum space requires the knowledge of transverse-momentum-dependent PDFs (TMDs), defined via TMD factorization\tcite{Collins:1981uk,Collins:2011zzd}.
Striking successes have been obtained in recent years in the quark-TMD sector. The deep knowledge on their formal properties has been corroborated by encouraging results at the level of phenomenology.
Conversely, the gluon-TMD sector still represents an almost uncharted territory. Spin-dependent gluon TMD distributions were classified in Refs.\tcite{Mulders:2000sh,Meissner:2007rx,Lorce:2013pza,Boer:2016xqr}. The first phenomenological studies appeared only recently \cite{Lu:2016vqu,Lansberg:2017dzg,Gutierrez-Reyes:2019rug,Scarpa:2019fol,Adolph:2017pgv,DAlesio:2017rzj,DAlesio:2018rnv,DAlesio:2019qpk}.
At variance with collinear PDFs, TMDs are not universal\tcite{Brodsky:2002cx,Collins:2002kn,Ji:2002aa}.
From a formal perspective, their process dependence is due to the structure of the Wilson lines.
Gluon TMDs have a more diversified structure of \emph{modified universality} than quark TMDs because different classes of reactions pick distinct gluon gauge-link structures. 
Two major kinds of gluon gauge links emerge: the $f\text{-type}$, also known as Weisz\"acker--Williams structure, and the $d\text{-type}$, known as dipole structure~\cite{Kharzeev:2003wz,Dominguez:2010xd,Dominguez:2011wm}.
At small $x$ and large transverse momentum, the unpolarized ($f_1^g$) and Boer--Mul\-ders ($h_1^{\perp^g}$) gluon TMDs are connected to the small-$x$ \emph{unintegrated gluon distribution} (UGD), whose evolution is regulated by the Balitsky--Fadin--Kuraev--Lipatov (BFKL) equation~\cite{Fadin:1975cb,Balitsky:1978ic} (see Refs.\tcite{Celiberto:2017ius,Celiberto:2020wpk,Nefedov:2021vvy,Hentschinski:2021lsh,Celiberto:2022fgx} for formal studies and Refs.\tcite{Hentschinski:2012kr,Besse:2013muy,Celiberto:2016vhn,Bolognino:2019yqj,Bolognino:2018rhb,Bolognino:2018mlw,Bolognino:2019bko,Bolognino:2019pba,Celiberto:2019slj,Bautista:2016xnp,Garcia:2019tne,Hentschinski:2020yfm,Bolognino:2021niq,Bolognino:2021gjm,Bolognino:2022uty,Celiberto:2022fam,Bolognino:2022ndh,Brzeminski:2016lwh,Celiberto:2018muu} for phenomenological applications).
The gluon Boer--Mul\-ders function carries information on linearly-polarized gluons in an unpolarized nucleon. It is responsible of spin effects in collisions of unpolarized hadrons~\cite{Boer:2010zf,Sun:2011iw,Boer:2011kf,Pisano:2013cya,Dunnen:2014eta,Lansberg:2017tlc}. Its weight is expected to become more and more relevant as $x$ becomes smaller.
The gluon Sivers TMD, $f_{1T}^{\perp g}$, describes unpolarized gluons in a transversely-polarized hadron. It allows us to access transverse-spin asymmetries rising in processes with polarized-proton beams.
The QCD-Odderon connection~\cite{Boussarie:2019vmk} makes it possible to study the gluon Sivers function also via unpolarized electron-nucleon scatterings.

Due to the lack of experimental data for gluon TMDs, exploratory analyses through simple and flexible models are needed. Pioneering studies were performed by making use of the so-called \emph{spectator framework}~\cite{Lu:2016vqu,Mulders:2000sh,Pereira-Resina-Rodrigues:2001eda}.
Originally adopted to model quark TMDs~\cite{Bacchetta:2008af,Bacchetta:2010si,Gamberg:2005ip,Gamberg:2007wm,Jakob:1997wg,Meissner:2007rx}, it assumes that a gluon is taken from the struck hadron describing the residual system 
as an effective on-shell spin-1/2 particle.
Leading-twist spectator-model T-even gluon TMD densities were recently calculated~\cite{Bacchetta:2020vty} (see also Refs.~\cite{Celiberto:2021zww,Bacchetta:2021oht,Celiberto:2022fam,Bacchetta:2022crh}), while results for the T-odd ones are at a preliminary level~\cite{Bacchetta:2021lvw,Bacchetta:2021twk,Bacchetta:2022esb}.

An intriguing perspective is offered by the study of quarkonium states in proton-proton and lepton-proton collisions. On the one hand, quarkonium detections at low transverse momentum (see, \emph{e.g.} Refs.\tcite{Kishore:2019fzb,DAlesio:2019gnu,Boer:2020bbd,DAlesio:2020eqo,Bacchetta:2018ivt,Boer:2021ehu,DAlesio:2021yws}) allow one to measure the gluon polarization because they are dominated by gluon-induced hard scatterings already at the Born level.
On the other hand, they are a clean probe of the $f$-type gluon gauge link\tcite{Boer:2016bfj}.
In this work we present preliminary results for the $\eta_{b,c}$ production in TMD factorization. 
By making use of kinematic cuts tailored on the current acceptances at LHCb and on the nominal ones at a future LHC fixed-target experiment\tcite{Brodsky:2012vg,Lansberg:2012kf,Feng:2015cba,Kikola:2017hnp,Hadjidakis:2018ifr,Trzeciak:2021dyb}, we highlight the potential of exploring the phenomenology of unpolarized and linearly-polarized gluon TMDs.

\section{$\eta_{b,c}$ pseudoscalars in $pp$ collisions}
\label{eta_Q}

\begin{figure*}[!t]

\centering

\includegraphics[width=0.45\textwidth]{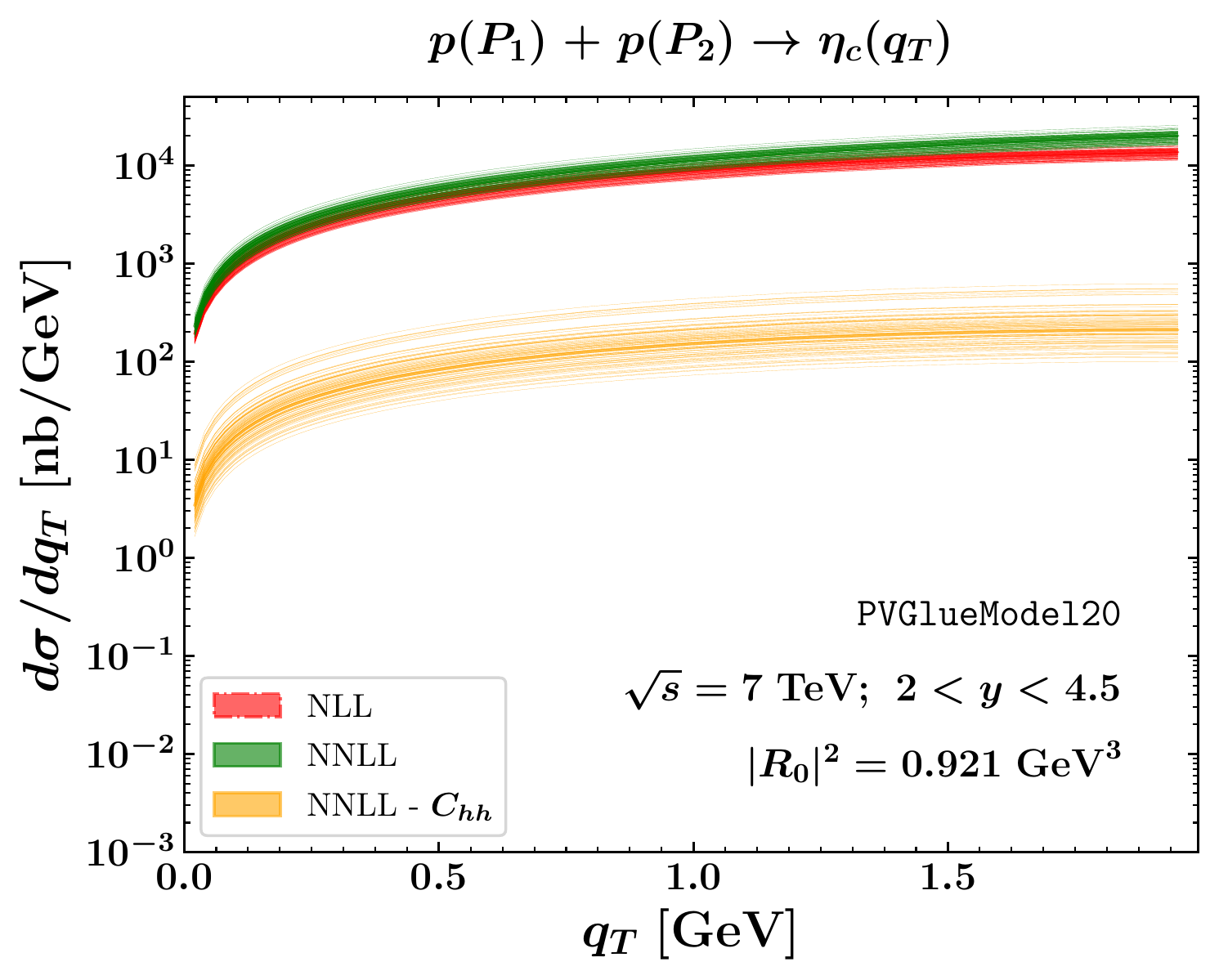}
\hspace{0.50cm}
\includegraphics[width=0.45\textwidth]
{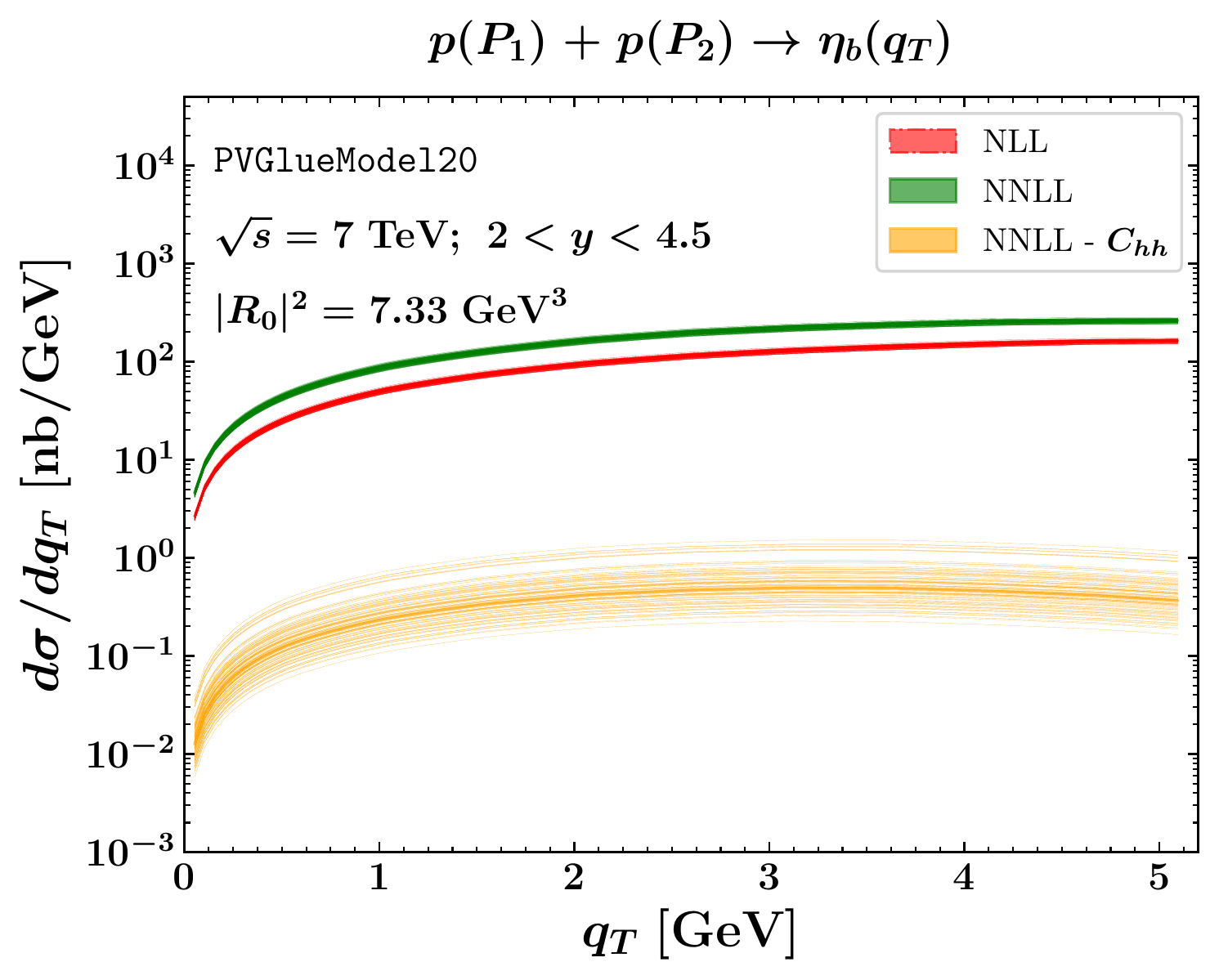}

\includegraphics[width=0.45\textwidth]
{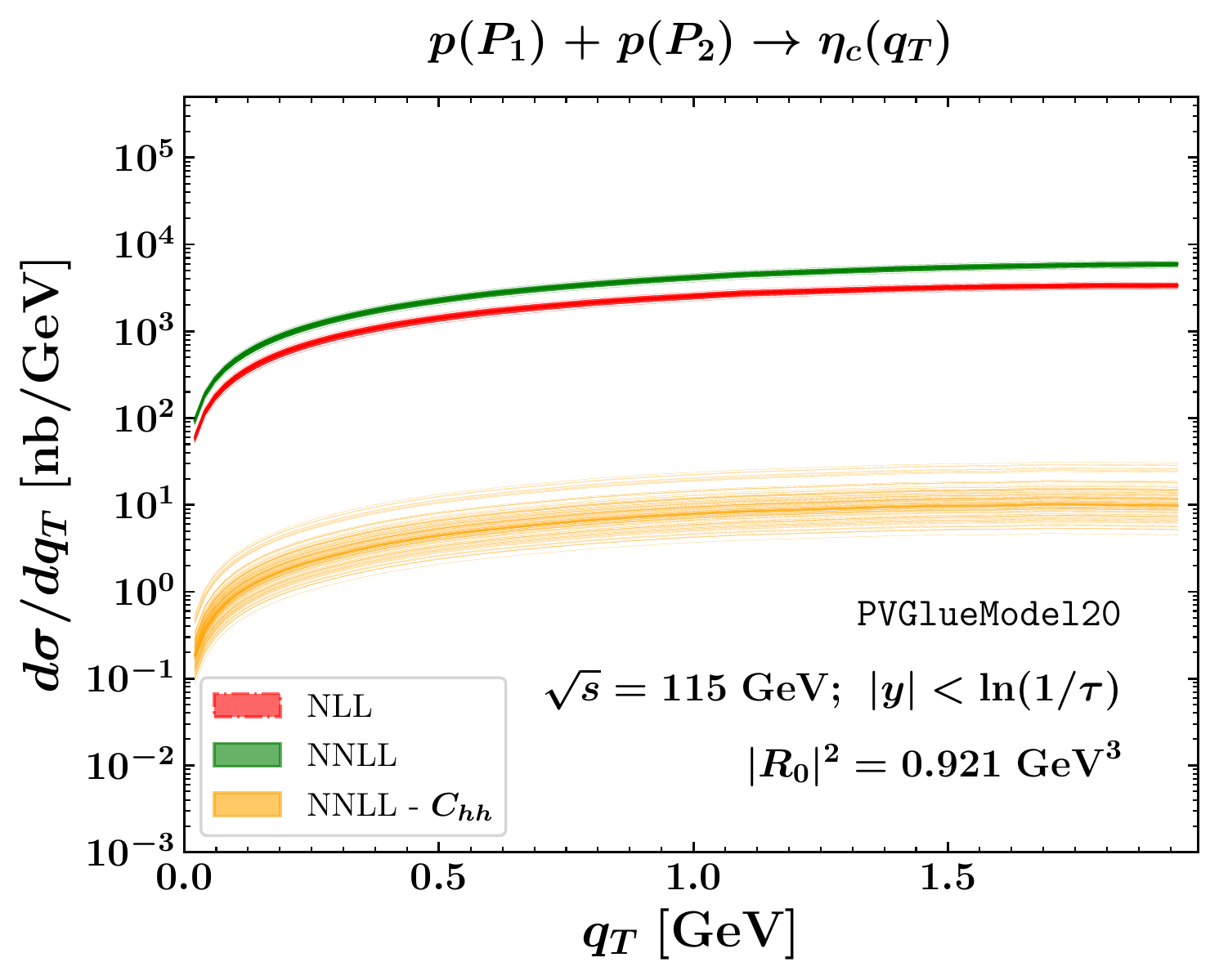}
\hspace{0.50cm}
\includegraphics[width=0.45\textwidth]
{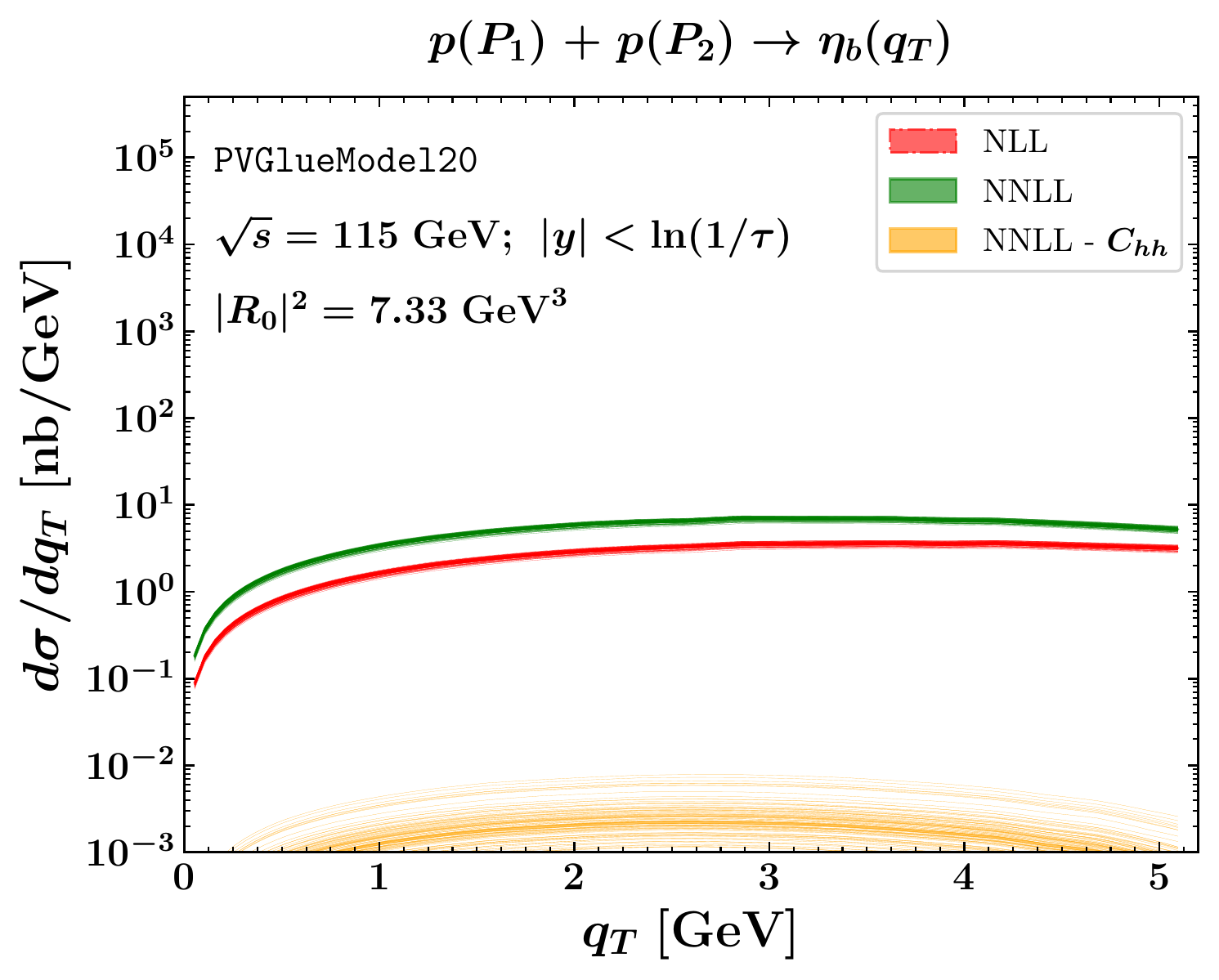}

\caption{Transverse-momentum distributions of $\eta_c$ (left) and $\eta_b$ (right) produced at 7 TeV at LHCb (upper) and at 115 GeV at AFTER (lower). Red and green bands for NLL and NNLL accuracy in the resummation of large logarithms from soft gluon radiation, respectively. Orange band for NNLL contribution of Boer--Mulders only. 
$R_0$ stands for the radial wave function at the origin of the quarkonium state\tcite{Butenschoen:2014dra,Maltoni:2004hv}.}
\label{fig:dsigma_dqT}
\end{figure*}

The dominant mechanism describing the production of a quark\-onium state at low transverse momentum is the \emph{short-distance} emission of a $(Q \bar Q)$ heavy-quark pair directly produced in the hard scattering, followed by the nonperturbative hadronization.
An effective description of such a mechanism is provided by the non-relativistic QCD (NRQCD) framework~\cite{Caswell:1985ui,Thacker:1990bm,Bodwin:1994jh}.
It is based on the assumption that all possible Fock states contribute to the emitted quarkonium through a linear superposition ($|Q \bar Q\rangle$, $|Q \bar Q g\rangle$, ...). All these factors are calculated in terms of a double expansion in powers of the strong coupling, $\alpha_s$, and of the relative velocity of the two heavy quarks, $v$.
Differential distributions sensitive to the emission of a quarkonium are given as a sum of partonic hard factors, each one corresponding to the production of a given Fock state and being multiplied by a long-distance matrix element (LDME) encoding information about the nonperturbative hadronization process.
LHCb data for $\eta_c$ production at moderate transverse momentum
are well described by the color-singlet configuration only\tcite{Butenschoen:2014dra}, and the same holds for $\eta_b$ mesons\tcite{Brodsky:2012vg,Lansberg:2012kf,Maltoni:2004hv}.
Therefore, in our study we build the differential cross section by combining the standard TMD formalism\tcite{Collins:1981uk,Collins:2011zzd} with the NRQCD color-singlet contribution for pseudoscalar mesons.
The leading-order TMD cross section, differential in the rapidity $y$ and transverse momentum $\boldsymbol{q}_T$ of the produced pseudoscalar meson, reads
\begin{eqnarray}
\label{eq:dsigma_LO}
 \frac{\drv \sigma}{\drv y \, \drv^2 \boldsymbol{q}_T} &=& \frac{2 \pi^3}{9} \frac{ \alpha_s^2}{m_{\eta_{c,b}}^3 s} \, {\cal H} \, \langle 0 | {\cal O}_1^{\eta_{c,b}} (^1S_0) | 0 \rangle
 \\ \nonumber
 &\times&
 \left\{ {\cal C} \, [f_1^g f_1^g] - {\cal C} \, [\omega_{UU} \, h_1^{\perp g} h_1^{\perp g}] \right\} \; ,
\end{eqnarray}
where $m_{\eta_c} = 2.9836$ GeV and $m_{\eta_b} = 9.398$ GeV are the quarkonium masses, $s$ is the center-of-mass energy, ${\cal H}$ is the hard function (see, \emph{e.g.}, Eq.~(14) of Ref.\tcite{Signori:2016jwo}) and
\begin{equation}
\label{eq:LDME}
 \langle 0 | {\cal O}_1^{\eta_{c,b}} (^1S_0) | 0 \rangle = \frac{N_c}{2 \pi} |R_0|^2 [1 + {\cal O}(v^4)]
\end{equation}
is the NRQCD matrix element for $\eta_{c,b,}$ produced in color-singlet configurations, with $N_c$ the QCD color number and $R_0$ the radial wave function at the origin\tcite{Butenschoen:2014dra,Maltoni:2004hv}.
The TMD convolution operator ${\cal C}$ and the transverse-momentum weight $\omega_{UU}$ in Eq.\eref{eq:dsigma_LO}, as well as details on the next-to-leading order expression for the cross section, can be found in Refs.\tcite{Signori:2016jwo,Signori:2016lvd}.
In our analysis we describe the unpolarized gluon TMD, $f_1^g$, and the Boer--Mulders one, $h_1^{g \perp}$, in terms of the {\tt PVGlueModel20} model of Ref.\tcite{Bacchetta:2020vty} at the initial scale. Their evolution is controlled by the standard Collins--Soper--Sterman (CSS) equations\tcite{Collins:1981uk,Collins:2011zzd}.

In Fig.\tref{fig:dsigma_dqT} we show the transverse-mo\-men\-tum distribution of $\eta_c$ (left panels) and $\eta_b$ (right panels) produced at $\sqrt{s} = 7$ TeV at LHCb (upper panels) and at $\sqrt{s} = 115$ GeV at a fixed-target LHC experiment, labeled as AFTER (lower panels). The rapidity of the detected quarkonium is in the range $2 < y < 4.5$, as constrained by LHCb acceptances, or in the $|y| < \ln (1/\tau)$ window, with $\tau = m_{\eta_{c,b}}/\sqrt{s}$, for the fixed-target case. Uncertainty bands are obtained by using all 100 replicas of the {\tt PVGlueModel20} model calculation of $f_1^g$ and $h_1^{\perp g}$ TMDs \cite{Bacchetta:2020vty}. 
Red and green bands correspond to next-to-leading-log (NLL) and next-to-next-to-leading-log (NNLL) accuracies in the resummation of large logarithms from soft gluon radiation, respectively. Orange bands are for the NNLL Boer--Mulders contribution only.

\section{Future perspectives}
\label{conclusions}

We presented preliminary results for the $\eta_{b,c}$ hadroproduction in TMD factorization.
By studying transverse-momentum distributions tailored on the current acceptances of LHCb and on the nominal ones of a future LHC fixed-target experiment, we highlighted the potential of accessing the phenomenology of unpolarized and linearly-polarized gluon TMDs.
Future analyses will extend this work to: $(i)$ exploring the connection to emissions of quarkonium states at large transverse momentum in proton-proton and lepton-proton collisions\tcite{Flore:2020jau,Lansberg:2020ejc,Lansberg:2021vie,Lansberg:2019adr}, $(ii)$ enhancing the description of quark\-onium production mechanisms in terms of \emph{shape functions}\tcite{Echevarria:2019ynx,Fleming:2019pzj}, $(iii)$ exploring common ground with similar studies in the high-energy regime via the recently discovered property of \emph{natural stability} of the high-energy resummation\tcite{Celiberto:2017nyx,Bolognino:2018oth,Bolognino:2019yls,Bolognino:2019ccd,Celiberto:2020tmb,Celiberto:2020rxb,Bolognino:2021mrc,Celiberto:2021dzy,Celiberto:2021fdp,Celiberto:2022dyf,Celiberto:2022kza,Celiberto:2022rfj,Celiberto:2022zdg,Celiberto:2022keu,Celiberto:2022gji}.

\section*{Acknowledgments}
\label{sec:acknowledgments}

We thank Jean-Philippe Lansberg, Melih A. Ozcelik and Miguel G. Echevarria for collaboration.
FGC thanks the Universit\`a degli Studi di Pavia for the warm hospitality.
AS was partly supported by the European Commission through the Marie Sklodowska-Curie Action SQuHadron (grant agreement ID: 795475).

\bibliographystyle{apsrev}
\bibliography{references}

\end{document}